\documentclass[twocolumn,pre,superscriptaddress,nofootinbib,aps,10pt]{revtex4-2}
\usepackage{amsfonts}
\usepackage{amsmath}
\usepackage{amssymb}
\usepackage{color}
\usepackage[toc,page]{appendix}
\usepackage{graphicx}
\usepackage{bm}
\usepackage{esint}
\usepackage{ulem}
\usepackage{xcolor}

\usepackage{soul}
\usepackage{cancel}
\usepackage{stmaryrd}

\begin{document}

\title{Coulomb Blockade in a One-Dimensional Array of Strong Josephson Junctions}

\date{\today}
\author{Bayan Karimi}
\affiliation{Pico group, QTF Centre of Excellence, Department of Applied Physics, Aalto University School of Science, P.O. Box 13500, 00076 Aalto, Finland}
\affiliation{Pritzker School of Molecular Engineering, University of Chicago, Chicago IL 60637, USA}
\author{Ze-Yan Chen}
\affiliation{Pico group, QTF Centre of Excellence, Department of Applied Physics, Aalto University School of Science, P.O. Box 13500, 00076 Aalto, Finland}
\author{Jukka P. Pekola}
\affiliation{Pico group, QTF Centre of Excellence, Department of Applied Physics, Aalto University School of Science, P.O. Box 13500, 00076 Aalto, Finland}
\author{T.~Yamamoto}
\affiliation{Univ. Grenoble Alpes, CEA, Grenoble INP, IRIG, PHELIQS, 38000 Grenoble, France}
\affiliation{Institute for Solid State Physics, the University of Tokyo, Kashiwa, Chiba 277-8581, Japan}
\author{M.~Houzet}
\affiliation{Univ. Grenoble Alpes, CEA, Grenoble INP, IRIG, PHELIQS, 38000 Grenoble, France}
\author{L.~I.~Glazman}
\affiliation{Department of Physics and Yale Quantum Institute, Yale University, New Haven, Connecticut 06520, USA}
\begin{abstract}
We measure current across a voltage-biased, long array of Josephson junctions. Current-voltage characteristics do not reveal any critical current and instead display the presence of threshold voltage and finite activation energy, consistent with the Coulomb blockade. However, the respective energy scale is well below the value prescribed by a single-island geometric capacitances. We explain it by strong quantum fluctuations of charge associated with a large value of Josephson energy, $E_J\gg E_C$, where $E_C$ is the charging energy associated with the inter-island capacitance. Comparison between the data and theory points to an exponential (in parameter $E_J/E_C$) suppression of the threshold. We achieve a quantitative agreement with theory by accounting for charge-$e$ solitons associated with the quasiparticles entering the array from its leads. This scenario however underestimates the current below threshold in comparison with the experiment.
\end{abstract}
\maketitle

The zero-temperature transition between the superconducting and insulating states of a one-dimensional (1D) chain of small superconducting islands coupled by Josephson junctions has long been considered as a paradigmatic example in the theory of quantum phase transitions~\cite{Sondhi1997}. Yet, numerous attempts to see the transition in the current-voltage characteristics~\cite{Haviland1996,Chow1998,Kuo2001,Anderson2003,Miyazaki2002,Duty2017,Vogt2015} fell short of identifying the transition in the strong Josephson coupling regime. In this work, we uncover the culprit behind this problem: a large value of the Josephson energy $E_J$ compared to the charging energy $E_C = e^2/2C$ associated with the inter-island capacitance $C$. While the ratio $E_J /E_C$ is irrelevant for the existence of the transition and for the critical value of the Josephson energy $E_J$, it may render the threshold bias in the insulating state unobservably small~\cite{Haviland2000}. 

It is well-known that the superconductor-to-insulator transition in such system is of Berezinskii-Kosterlitz-Thouless type and is controlled by the dimensionless ratio $E_J /E_g$, where $E_g = e^2/2C_g$ is the charging energy associated with the stray capacitance $C_g$ of an island. At $C\gg C_g$, the critical value of that ratio is approximately  $32/\pi^2$~\cite{Bradley1984,Korshunov1989,Sondhi1997,Basko2020}. At below-critical values $E_J/E_g$, and regardless the value of $E_J /E_C$, the array enters an insulating state, with a finite bias threshold for the charge transport along it. Above the threshold, charge propagates along the array in the form of solitons, each carrying charge $2e$. The energy of a static soliton defines the characteristic scale for transport activation energy and threshold bias~\cite{Haviland2000}. A common lore is that the spatial extent of a soliton is $\propto \sqrt{C/C_g}$ and, respectively, the soliton energy is $\sim e^2/\sqrt{C C_g}$ at $C\gg C_g$. We note, however, that these estimates are valid only in the limit $E_C\gg E_J$~\cite{BenJacob1989,Bakhvalov1989,Averin1991,Hermon1996,Haviland1996}. In the opposite limit, the electric polarization associated with the transfer of a Cooper pair from one island to another is controlled not by $C$ but, instead, by the quantum capacitance $C_\star = e^2/\pi^2 \lambda$~\cite{Likharev1985,Houzet2024}, where
\begin{equation}
\label{eq:lambda}
\lambda= \frac 8{\sqrt{\pi}} (8 E_J^3 E_C)^{1/4}e^{-\sqrt{8 E_J/E_C}}\quad\mathrm{at}\quad E_J\gg E_C
\end{equation}
is the phase-slip amplitude of a Josephson junction in the array. Replacement $C \to  C_\star$ results in an exponentially small soliton energy, $\sim e\sqrt{\lambda/C_g}$, and suppressed values of threshold voltage and activation energy. In the experimental realizations, the stray capacitance $C_g$ is by far smaller than the capacitance $C$ of the tunnel barrier comprising the Josephson junction. Therefore, the ratio $E_J /E_C$ remains large even deep in the insulating phase, at $E_J /E_g$ substantially smaller than its critical value $\sim 32/\pi^2$. The small soliton energy hinders the observation of the insulating phase in transport measurements.

In this work, we report a successful observation of the Coulomb blockade in a nominally-insulating ($E_g\gg E_J$) long array of Josephson junctions, despite the exponentially strong suppression of the threshold voltage caused by the quantum charge fluctuations at $E_J\gg E_C$. In parallel, we develop a quantitative theory for the threshold voltage. It associates the threshold with the bias needed to ``push in'' a soliton into and across the array. For a uniform array, we find that the theoretical value of the threshold associated with an extra Cooper pair charge-$2e$ exceeds the measured value by a factor of two; the theory also overestimates the activation energy by a factor of four. Random displacement charges, a rather ubiquitous kind of disorder, ``softens'' the energy gap and replaces it by so-called Larkin energy. That could have explained a reduced threshold bias for a short array, but not for a relatively long one studied in our experiment. These observations prompt us to consider the role of quasiparticles, ubiquitous in the leads. Quasiparticles enter the array as charge-$e$ solitons. The corresponding theoretical values of the threshold bias and activation energy are in excellent agreement with the experiment. However, thermally-activated motion of charge-$e$ solitons along the array results in a current which is much lower than that observed in the experiment. This discrepancy is currently unresolved.

\vspace{0.25in}

\begin{figure}
\includegraphics[width=1.0\columnwidth]
{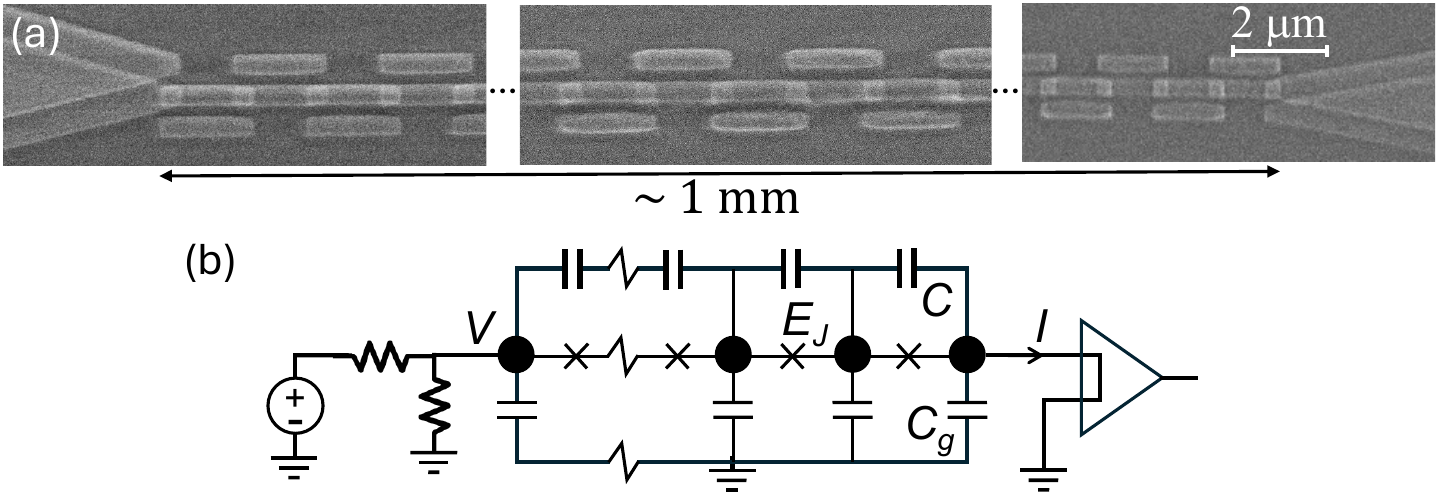}
\caption{\label{F:1}
Layout of the Josephson junction array studied in this work. (a) Image of the circuit fabricated by angle evaporation. (b) The circuit that represents the array consists of superconducting islands that are connected in series by junctions with Josephson energy $E_J$ and capacitance $C$. Furthermore, each island has stray capacitance~$C_g$. The measurement setup is also represented.
}
\end{figure} 
In the experiment, we first characterize the parameters of the studied array of Josephson junctions. It consists of $N=825$ Al/Al$_2$O$_3$/Al junctions in series, see Fig.~\ref{F:1}. The resistance of the array in the normal state is $R_{N}\approx 8.96$~M$\Omega$, yielding the normal state resistance $R=R_{N}/N\approx 10.9$~k$\Omega$ per junction. The Josephson energy $E_J/k_B\approx 0.684$~K ($E_J\approx59~\mu e$V) was then estimated using the Ambegaokar-Baratoff formula for the critical current, $I_c=\pi\Delta/2eR$, the superconducting gap $\Delta_{\rm Al}\approx200~\mu e$V for aluminium film with thickness 50~nm, and the found value of $R$. Coulomb blockade thermometry~\cite{Pekola1994} at $T\approx4.2$~K was used to estimate the charging energy $E_C/k_B\approx 0.226$~K ($E_C\approx20~\mu e$V). Then Eq.~\eqref{eq:lambda} gives the estimate $\lambda/k_B\approx 29$~mK ($\lambda \approx 2.5~\mu e$V), while a more accurate estimate relying on the dispersion of the energy levels in a Cooper pair box~\cite{Koch2007} yields $\lambda/k_B\approx 23$~mK ($\lambda \approx 2.0~\mu e$V).  Furthermore, we estimate the stray capacitance along the lines of Refs.~\cite{Wahlgren1995,Kauppinen1996} and find $C_g\approx 50~$aF, corresponding to $E_g/k_B\approx18.6$~K ($E_g\approx1.6$~m$e$V). (Here we used $C_g\simeq2\pi \varepsilon_0\varepsilon_{\rm eff} l/\ln(8 h/w)$ for a thin Al island of length $l\approx1.5~\mu{\rm m}$ and width $w\approx0.2~\mu{\rm m}$, and $\varepsilon_{\rm eff}=(1+\varepsilon_r)/2$ with relative permittivity $\varepsilon_r\approx11.9$ of the Si substrate at a distance $h=500~\mu$m from the island.) These numbers yield $E_J/E_g\approx 0.04$, clearly placing the array deep in the insulating regime.
\begin{figure*}
\includegraphics[width=1.0\textwidth]
{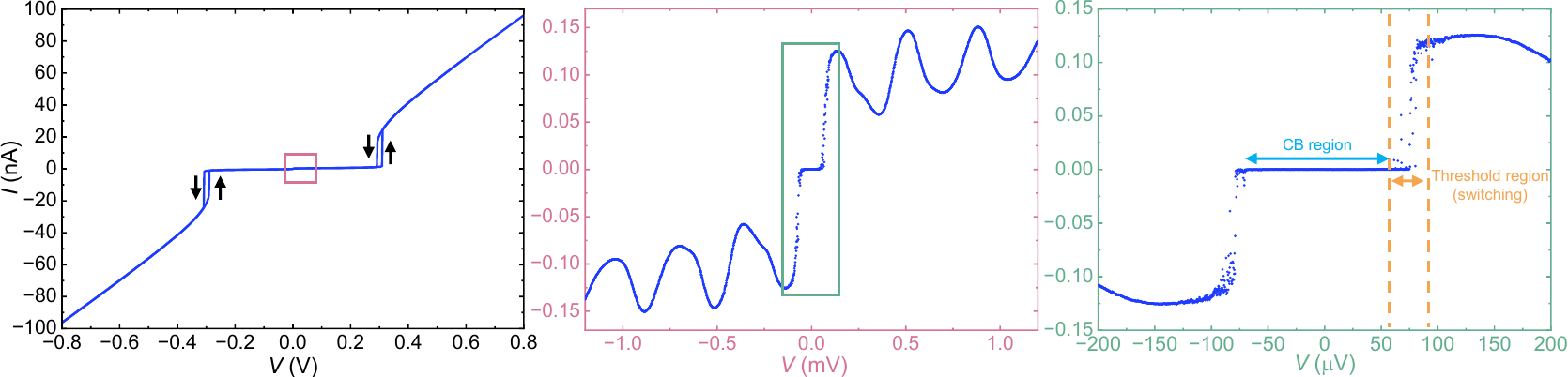}
\caption{\label{F:2}
Current-voltage characteristics of a dc voltage-biased Josephson junction array on three increasingly fine scales of bias (left to right) at base temperature $T\approx42.9$~mK. The Coulomb blockade (CB) threshold is clearly seen on the finest-scale plot.
}
\end{figure*} 

The current-voltage characteristic measured across the Josephson junction array when a dc voltage bias is applied to one of its ends, while the other one is grounded, is presented in Fig.~\ref{F:2}. On the most coarse voltage scale, see the left panel, the array appears to display a finite critical current. Zooming in the vicinity of the apparent current jump at $V=0$, we see, however, current oscillations occurring on a finer scale comparable to the bias that breaches the superconducting gap in Al (central panel of Fig.~\ref{F:2}). We attribute these oscillations to Josephson junctions switching one by one into their normal state, each time the voltage increases by $\sim 2\Delta_{\rm Al}/e$~\cite{Haviland2000,Dolata2005,Crescini2023,Mukhopadhyay2023}.

The most spectacular observation emerges once we further zoom into the vicinity of zero bias. There, we find a clear evidence of the Coulomb blockade of the current across the array, with the threshold voltage in the range of $\sim 60-90~\mu$V (right panel of Fig.~\ref{F:2}). This range is significantly below the estimate~\cite{Bakhvalov1989} $e/(2\sqrt{CC_g})\approx 177~\mu$V based on the values of the geometrical capacitances.

\begin{figure*}
\includegraphics[width=1.0\textwidth]
{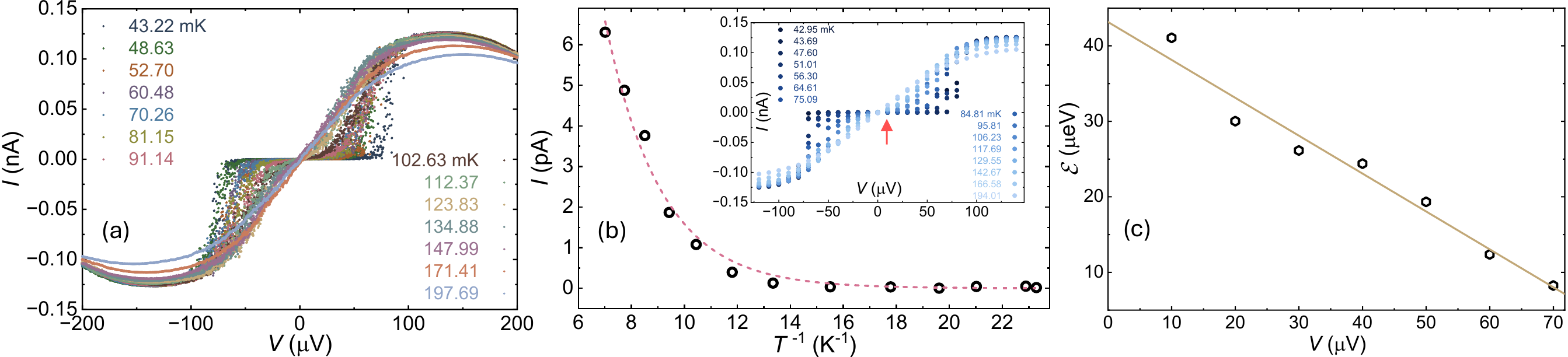}
\caption{\label{F:3}
(a) Temperature dependence of the current-voltage characteristics of a dc voltage-biased Josephson junction array. (b) Fit of the temperature dependence of the current measured at $V=10~\mu$V (indicated by the red arrow in the inset) with an Arrhenius law, which allows extracting the activation energy. (c) Voltage dependence of  the activation energy. Linear extrapolation (solid line) is used to estimate the threshold voltage.}
\end{figure*}

To further refine the determination of the threshold voltage, we measure the thermally-activated electron transport at biases below its threshold value. Figure~\ref{F:3}(a) shows the results of the temperature dependence of current at a set of biases spanning an interval $[-200~\mu {\rm V}; 200~\mu{\rm V}]$. We find that the temperature dependence is well described by the Arrhenius law, $I(V)\propto \exp[-{\cal E}(V)/k_BT]$, with activation energy ${\cal E}(V)$ depending on the bias $V$. An example of the Arrhenius plot for the current at the lowest bias, $V\approx10~\mu$V, is shown in Fig.~\ref{F:3}(b). The corresponding activation energy is ${\cal E}_0/k_B\approx 0.48$~K (${\cal E}_0\approx 42~\mu e$V). Additional plots obtained for other biases, $V = 10\text{–}70~\mu$V, are relegated to the Supplemental Material (SM). The activation energies extracted from all the plots are shown, as a function of bias in Fig.~\ref{F:3}(c).

To summarize our experimental findings: First, we see clear evidence of the Coulomb blockade of the current through the array; the threshold bias determined at the base temperature is in the range of $\sim 60-90~\mu$V, see right panel of Fig.~\ref{F:2}. Second, below the threshold, the activation energy of the current ${\cal E}(V)$ is a decreasing function of bias $V$, starting with ${\cal E}(10~\mu{\rm V})\approx 42~\mu e$V, see Fig.~\ref{F:3}(c). Third, as the notion of threshold voltage is defined at $T=0$, we refine its experimentally found value by extrapolating ${\cal E}(V)$ to ${\cal E}\to 0$ leading to $V_{\rm th}\approx 87~\mu$V. 

\vspace{0.25in}

We start the theoretical analysis with finding the threshold bias $V_{\rm th}$ and activation energy ${\cal E}(V)$ for an ideal array, in the absence of stray charges. For this we consider an array with finite length $d=Na$, where $a$ is the unit cell length; we assume that an external bias voltage $V$ is applied to the array's left end, while the right end is grounded. We introduce the charge displacement variable $q_n$, so that the charge in island $n$ is
\begin{equation}
\label{eq:deltaQ}
q_n-q_{n+1} = C_gV_n,
\end{equation}
where $V_n$ is the potential of the island with respect to the ground. Therefore, the voltage drop
across a junction, i.e., its (dynamical) electric polarization is $V_n-V_{n+1}$. In the quasi-static limit (typical frequencies much smaller than the Josephson plasma frequency $\omega_0=\sqrt{8E_JE_C}/\hbar$) and under the condition of a small phase-slip amplitude, $\lambda\ll \hbar \omega_0$, the relation between the polarization and charge displacement takes the form
\begin{equation}
\label{eq:deltaV}
V_n -V_{n+1} = L\frac{d^2q_n}{dt^2}+\frac{\pi \lambda}e\sin\frac{\pi q_n}e,
\end{equation}
where $L=\hbar^2/4e^2 E_J$ is the Josephson inductance.
Combining Eqs.~\eqref{eq:deltaQ} and \eqref{eq:deltaV}, and taking the continuous limit (justified in the case of a charge soliton being much longer than the unit cell length), we find the following closed equation for the charge displacement field $q(x,t)$:
\begin{equation}
\label{eq:EOM}
\frac{\partial^2q}{\partial x^2}=LC_g\frac{\partial^2q}{\partial t^2} +\frac{\pi \lambda C_g}e\sin\left(\frac{\pi q}e\right).
\end{equation}
Here the length is measured in units of the elementary cell size of the array. The
current flowing in the array is expressed as
\begin{equation}
\label{eq:I}
I(x,t) =\frac{\partial }{\partial t}q(x,t).
\end{equation}
We then express the boundary conditions for Eq.~\eqref{eq:EOM} using Eq.~\eqref{eq:deltaQ}. The bias voltage $V$ applied at the array's left end ($x = 0$) leads to the condition
\begin{equation}
\label{eq:BCLeft}
V = -\frac1{C_g}\left.\frac{\partial q}{\partial x}\right|_{x=0},
\end{equation}
while the zero-voltage condition at the right end of the array reads
\begin{equation}
\label{eq:BCRight}
0=-\frac{1}{C_g}\left. \partial_x q\right|_{x=d}.
\end{equation}
The threshold voltage may then be identified as the highest bias voltage that allows for a stationary solution ($\partial_t q=0$) of Eqs.~\eqref{eq:EOM}, \eqref{eq:BCLeft}, and \eqref{eq:BCRight}. For an array's length much longer than a charge soliton, the boundary condition~\eqref{eq:BCRight} may be substituted with $q(x \to \infty)=0$ for a half-infinite array. The first integral of the dynamic equation \eqref{eq:EOM} at $\partial_t q=0$, written for the point $x = 0$, and accounting for the boundary condition \eqref{eq:BCLeft}, then reads:
\begin{equation}
\label{eq:IM}
(C_gV)^2=2C_g\lambda \left(1-\cos\frac{\pi q(0)}{e}\right).
\end{equation}
A stationary value $q(0)$ may thus be found if the bias is below the threshold:
\begin{equation}
V_{\rm th} = 2\sqrt{{\lambda}/{C_g}}.
\label{eq:Vth}
\end{equation}
The threshold corresponds to the margin of static equilibrium: the charge displacement $q(0)$ caused by the bias reaches $e$ (half a Cooper pair charge).

Below threshold, an electrostatic barrier needs to be overcome for generating a charge soliton in the array. To determine that barrier, we note that the charge displacement field in the array at equilibrium can be obtained by extremizing the energy functional:
\begin{equation}
\label{eq:energy-functional}
{\cal E}[q(x)]=\int_0^\infty dx\left[\frac1{2C_g}(\partial_x q)^2+\lambda\left(1-\cos\frac{\pi q}e\right)\right]-Vq(0).
\end{equation}
By inserting the first integral of motion with $q(x \to \infty)=0$ into Eq.~\eqref{eq:energy-functional}, we find that it reduces to an energy functional on variable $q(0)$ (such that $|q(0)|<2e$),
\begin{equation}
\label{eq:energy-functional2}
{\cal E}[q(0)]=\frac{4e}\pi \sqrt{\frac\lambda{C_g}}\left[1-\cos\frac{\pi q(0)}{2e}\right]-Vq(0).
\end{equation}
The activation energy is then identified as the difference between maximal and minima energies \eqref{eq:energy-functional2} evaluated at two solutions $q(0)$ of Eq.~\eqref{eq:IM},
\begin{eqnarray}
\label{eq:activation}
{\cal E}(V)&=&\frac{4eV_{\rm th}}\pi f(V/V_{\rm th}),\qquad \\
f(x)&=&\sqrt{1-x^2}-x(\pi/2- \arcsin x).
\nonumber
\end{eqnarray} 
At zero voltage, the activation energy matches the soliton energy. Finite bias reduces the energy linearly in $2eV$ at $V\ll V_{\rm th}$; its suppression is $\propto (V_{\rm th}-V)^{3/2}$ near threshold.

\vspace{0.25in}

Let us now compare experiment and theory. With the quoted values for $E_J,E_C,E_g$, the soliton length $\ell_{\rm s}=\sqrt{2E_g/\pi^2\lambda}\approx 12$ (in units of $a$, i.e., encompassing about 12 superconducting islands) is much shorter than the array length and therefore well within the range of validity of our theory worked out for a half-infinite array. We note also, that the soliton is long on the scale of the elementary cell of the periodic array.

Using the array's parameters, we find $V_{\rm th}\approx 158~\mu$V for the threshold and ${\cal E} (0)\approx 202~\mu e$V for the activation energy at zero bias. These two estimates are systematically higher than the measured values. Specifically, the theoretical value of the threshold value evaluated according to Eq.~(\ref{eq:Vth}) exceeds the measured value by a factor $\sim 2$. The theoretical value of the activation energy ${\cal E}(0)$ exceeds the measured one approximately by a factor of $4$. The found theoretically bias dependence of the activation energy looks similar to the data. However, the overall slope characterized by the ratio ${\cal E} (0)/eV_{\rm th}$ is different: it has a universal value $4/\pi\simeq 1.3$ in theory, and an approximately twice smaller value ${\cal E} (0)/eV_{\rm th}\approx 0.6$ in the experiment.

\begin{figure}
\includegraphics[width=\columnwidth]{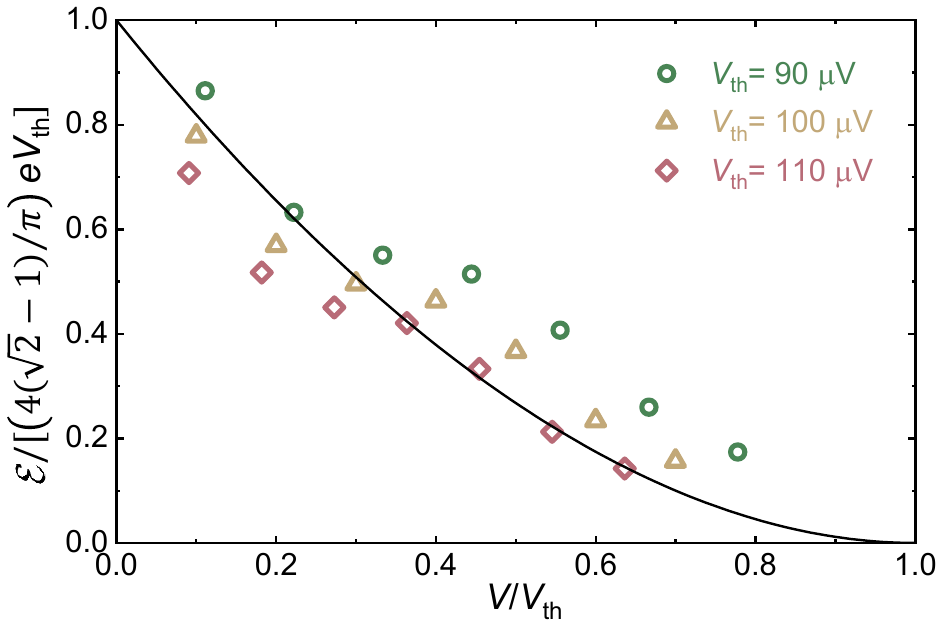}
\caption{\label{F:4}
Prediction for the voltage dependence of activation energy, Eq.~\eqref{eq:act-th}, associated with excess quasiparticles in the leads entering an ideal Josephson junction array to which they are connected. The experimental data points of Fig.~\ref{F:3}(c) are represented in dimensionless units by using three different values of the threshold voltage $V_{\rm th}$ as a single fit parameter.}
\end{figure}

At present, we do not have a definitive explanation of the systematic overestimate of the energy scales. Artificially, one may attribute it to a uniform charge displacement. Assuming the static charge displacement $Q$ per island, one finds ${\cal E}(0)=(4eV_{\rm th}/\pi)f(C_g|Q|/V_{\rm th})$. If $|Q|$ for an uniformly distributed charge exceeds the critical value $V_{\rm th}/C_g$, solitons proliferate spontaneously across the array, suppressing the activated behavior. However, a more realistic assumption of a maximal charge disorder has a nuanced effect on the energy scales. For a given static configuration of charge disorder, a collective pinning mechanism results in the localization of the charge density wave described (in the continuous-field approximation) by the charge displacement field $q(x)$. To estimate the localization length $\ell_\star$ (in units of $a$), one compares the elastic energy associated with of a Cooper pair charge accumulated over that length, $E_g/\ell_\star$, with the typical potential energy $\lambda\sqrt{\ell_\star}$, yielding the so-called Larkin length $\ell_\star = (E_g/\lambda)^{2/3}$. The corresponding value of the Larkin energy is $\omega_\star=\lambda\sqrt{l_\star}=(E_g\lambda^2)^{1/3}$. We estimate then ${\cal E}(0)\sim eV_{\rm th}\sim \omega_\star$. The quoted parameters for $E_J,E_C,E_g$ yield $\ell_\star\approx 87$ and $\omega_\star\approx 18~\mu e$V. According to the numerical results of \cite{Vogt2016}, the {\sl averaged over the disorder configurations} threshold voltage in a short array, $N\sim l_\star$, would be of the order of $\omega_\star/e$. However, the threshold, defined as an configuration-averaged value, grows with the array length, $eV_{\rm th}\approx 2.2 \omega_\star (N/\ell_\star)$, and exceeds the clean-limit value at $N \approx 1.3 (\ell_\star)^{5/4}\approx 340$. This crossover value is by far shorter than the measured array length. Our numerical simulations~\cite{Tsuyoshi}, which builds on efficient methods~\cite{Rosso2001,Kolton2009} that have been developed to determine the threshold voltage and activation energy in presence of an electric field in disordered elastic media, generally confirm the results of \cite{Vogt2016} for the average threshold values, fixing numerical prefactors in above formulas and yielding an average threshold voltage $V_{\rm th}\approx 382~\mu $V with the quoted array's parameters. However, we uncover a substantial width of the threshold distribution function, characterized by standard deviation $e\delta V_{\rm th}\approx 1.4 \omega_\star \sqrt{N/\ell_\star}\approx 78~\mu e$V. 

We expect the configuration averaging in the experiment happens because the random charge configurations vary on a time scale comparable to the measurement time. We have little information regarding the presence of charge disorder in our case. We may conjecture that in case disorder is present, the current measurement setup would preferentially select the ``distribution tail'' configurations corresponding to a lower threshold.

As an alternative explanation, we consider the effect of excess quasiparticles in the leads, which are inevitably present under the conditions of the experiment, i.e. with bare superconducting aluminum leads without quasiparticle traps~\cite{Saira2012}. The presence of an electron in an array's island leads to an $e$-jump in charge displacement field $q$ at the island's position $x$. Examining Eq.~\eqref{eq:IM} in presence of an $e$-jump at the array's edge, one then deduces that the maximal voltage at which a stationary charge displacement field is created in the array is reduced from $V_{\rm th}=2\sqrt{\lambda/C_g}$ to $\tilde V_{\rm th}=\sqrt{2\lambda/C_g}$, corresponding to the charge displacement caused by bias reaching  $q(0)=e/2$. 

To estimate the activation energy below threshold, $V<\tilde V_{\rm th}$, we insert a soliton profile with boundary condition $q(x\to \infty)=0$, as well as an $e$-jump from $e/2$ to $3e/2$ at position $x$, inside the energy functional~\eqref{eq:energy-functional}, to find 
\begin{equation} 
\label{eq:energy-e-jump}
{\cal E}[q(0)]=\frac {4e}\pi\sqrt{\frac\lambda{C_g}}\left[1-\sqrt{2}-\cos\frac{\pi q(0)}{2e}\right]-V(q(0)-e)
\end{equation}
for $3e/2<q(0)<2e$. Note that the location $x$ of the $e$-jump is uniquely related with $q(0)$ through the first integral of the dynamic equation~\eqref{eq:EOM} at $\partial_t q=0$, yielding 
\begin{equation}
\label{eq:soliton-position}
\tan\frac{\pi q(0)}{4e}=e^{x/\ell_{\rm s}}\tan\frac{3\pi}8.
\end{equation} 
Using Eq.~\eqref{eq:energy-e-jump} instead of Eq.~\eqref{eq:energy-functional2} in finding the energy maximized over the values of $q(0)$ allows us to find the activation energy in presence of an $e$-jump,
\begin{eqnarray} 
\label{eq:act-th}
\tilde {\cal E}(V)&=&\frac {4(\sqrt{2}-1)}\pi e\tilde V_{\rm th}\tilde f(V/\tilde V_{\rm th}),\\
\tilde f(x)&=&(\sqrt{2}+1)\left(\sqrt{2 - x^2} -1 +  x \arcsin\frac x{\sqrt{2}}-\frac{\pi x} 4 \right).
\nonumber \\
\end{eqnarray}
The voltage dependence of $\tilde {\cal E}(V)$ is shown in Fig.~\ref{F:4}. At zero voltage, the activation energy matches the soliton energy in presence of an $e$-jump. Finite bias reduces the energy linearly in $eV$ at $V\ll V_{\rm th}$; the activation energy  suppression is $\propto (\tilde V_{\rm th}-V)^2$ near threshold. 

Note that charge-$e$ soliton transport instead of the one produced by $2e$-solitons reduces the universal ratio ${\cal E}(0)/eV_{\rm th}$ from ${\cal E}(0)/eV_{\rm th}=4/\pi\approx 1.3$ to $\tilde {\cal E}(0)/e\tilde V_{\rm th}=4(\sqrt{2}-1)/\pi\approx 0.53$. The latter value is much closer to the one found in the experiment, ${\cal E}(0)/eV_{\rm th}\approx 0.6$. For the parameters of the array, we find the new theoretical  threshold voltage value $\tilde V_{\rm th}\approx 113~\mu$V; it is also closer to the experimentally found one, $V_{\rm th}\approx87~\mu$V. 

Next, we analyze the dependence of the activation energy on bias, ${\cal E}(V)$. For that we fit the data of Fig.~\ref{F:3}(c) by using Eq.~\eqref{eq:act-th} with $V_{\rm th}$ as a single fit parameter, see Fig.~\ref{F:4}. The admissible values of the fit parameter, $V_{\rm th}\approx 90\div 110~\mu$V, are within the range between the measured and theoretically estimated values of the threshold.

Our theory of $\tilde V_{\rm th}$ and ${\cal E}(V)$ assumes a dilute gas of charge-$e$ solitons. That requires a fairly low quasiparticle density, so that the characteristic distance between the islands ``poisoned'' by quasiparticles exceeds the soliton length $\ell_{\rm s}$. That sets a limit, $n_{\rm qp}\ll ({\cal V}\ell_{\rm s})^{-1}$, for the quasiparticle density $n_{\rm qp}$; here ${\cal V}$ is the volume of a single island. It is common to express $n_{\rm qp}$ in terms of a dimensionless parameter, $x_{\rm qp}= n_{\rm qp}/(2\nu \Delta)$, where $\nu$ is the normal-state density of states per spin. We re-write the condition for the quasiparticle density as $x_{\rm qp}\ll \delta/(2\Delta\ell_{\rm s})$, with $\delta=1/\nu {\cal V}$ being the average normal-state level spacing in an island. In thermal equilibrium, $x_{\rm qp}\simeq \sqrt{2\pi k_BT/\Delta}e^{-\Delta/k_BT}$ at $T\ll \Delta/k_B$. Taking $\nu\approx 1.76\times 10^{10}~e$V$^{-1}\mu$m$^{-3}$ in Al and island volume ${\cal V}\approx 1.5\times 0.2\times 0.1~\mu$m$^3$, we find that the inter-soliton distance exceeds their size at at temperatures below~$200$~mK -- a necessary condition for Eq.~\eqref{eq:act-th} to hold.

Current $I(V)\sim I_0 e^{-{\cal E}(V)/k_BT}$ produced by thermally-activated transport of charge-$e$ solitons is associated with single quasiparticles hopping between the islands. That sets the limit~\cite{Cole2015} on the pre-exponential factor, $I_0\sim \delta/eR$, smaller than in the experiment by three orders of magnitude. We do not know yet how to resolve the inconsistency.

\vspace{0.25in}

In conclusion, the charge transport across a one-dimensional array of strong Josephson junctions in our experiment displays clear signatures of the Coulomb blockade behavior. The corresponding activation energy of linear conductance and the threshold bias for nonlinear transport are well below the energy needed to break a Cooper pair. Therefore, we attribute the blockaded transport to the propagation of charge solitons along the array. In the absence of disorder, the threshold bias is associated with the soliton energy. In the presence of disorder, the charge transport is limited by the pinning of a charge density wave in the array. The ensemble-averaged value of the threshold bias in a long disordered array is predicted to exceed its ``clean limit" counterpart in the absence of disorder. The clean limit theory for charge-$2e$ solitons comes close to explaining the experiment but overestimates the threshold voltage and activation energy by a factor of $2\div 4$. The clean limit theory for charge-$e$ solitons yields quantitative agreement with experimentally measured threshold voltage and activation energy. However, it severely underestimates the sub-threshold current.

Our findings motivate several perspectives for future research. On the experimental side, a detailed analysis of charge transport in arrays of various lengths may help contrasting the difference between depinning theories in presence of an electric field (in the bulk) or potential (at an edge). Furthermore, the analysis of low-frequency noise in charge transport measurements may provide information on the slow dynamics of stray charges in the array's environment. On the theory side, in finding the threshold voltage we analyzed the onset of instability of a static configuration. The depinning of a charge density wave, however, is an intrinsic non-equilibrium problem. Therefore, a dynamic theory will be needed to predict the current-voltage characteristic and resolve current inconsistencies between theory and the experiment.

\vspace{0.25in}
We thank Daniil Antonenko for discussions and Joonas Peltonen for technical support. The Work at Aalto University received funding from the Research Council of Finland Centre of Excellence programme grant 336810 and grant 349601 (THEPOW) and QuantERA II Programme that has received funding from the EU’s H2020 research and innovation programme under the GA No 101017733. BK acknowledges funding from the European Union’s Research and Innovation Programme, Horizon Europe, under the Marie Sklodowska-Curie Grant Agreement No. 101150440 (TcQTD). The Work at Yale University was sponsored by Army Research Office (ARO) under grant no. W911NF-23-1-0051, and by the U.S. Department of Energy (DoE) grant no. DE-SC0012704, Office of Science, National Quantum Information Science Research Centers, Co-design Center for Quantum Advantage (C2QA 2.0). MH acknowledges funding from the Plan France 2030 through the projects NISQ2LSQ ANR-22-PETQ-0006 and FERBO ANR-23-CE47-0004 and thanks Yale Quantum Institute for hospitality. TY was supported by a JSPS Overseas Research Fellowship. We thank the facilities and technical support of Otaniemi Research Infrastructure for Micro and Nanotechnologies (OtaNano).

\onecolumngrid
\setcounter{section}{0}
\renewcommand{\thesection}{S\arabic{section}}
\setcounter{figure}{0} 
\renewcommand{\thefigure}{S\arabic{figure}} 
\renewcommand{\figurename}{Fig.} 
\setcounter{table}{0}
\renewcommand{\thetable}{S\arabic{table}}%
\setcounter{equation}{0} 
\renewcommand{\theequation}{S\arabic{equation}} 
\section*{Supplemental Material}
Here, we present plots at various bias voltages to extract the activation energy shown in Fig.~3\,(c) in the main text.

\begin{figure}[h!]
	\centering
	\includegraphics [width=0.58\textwidth] {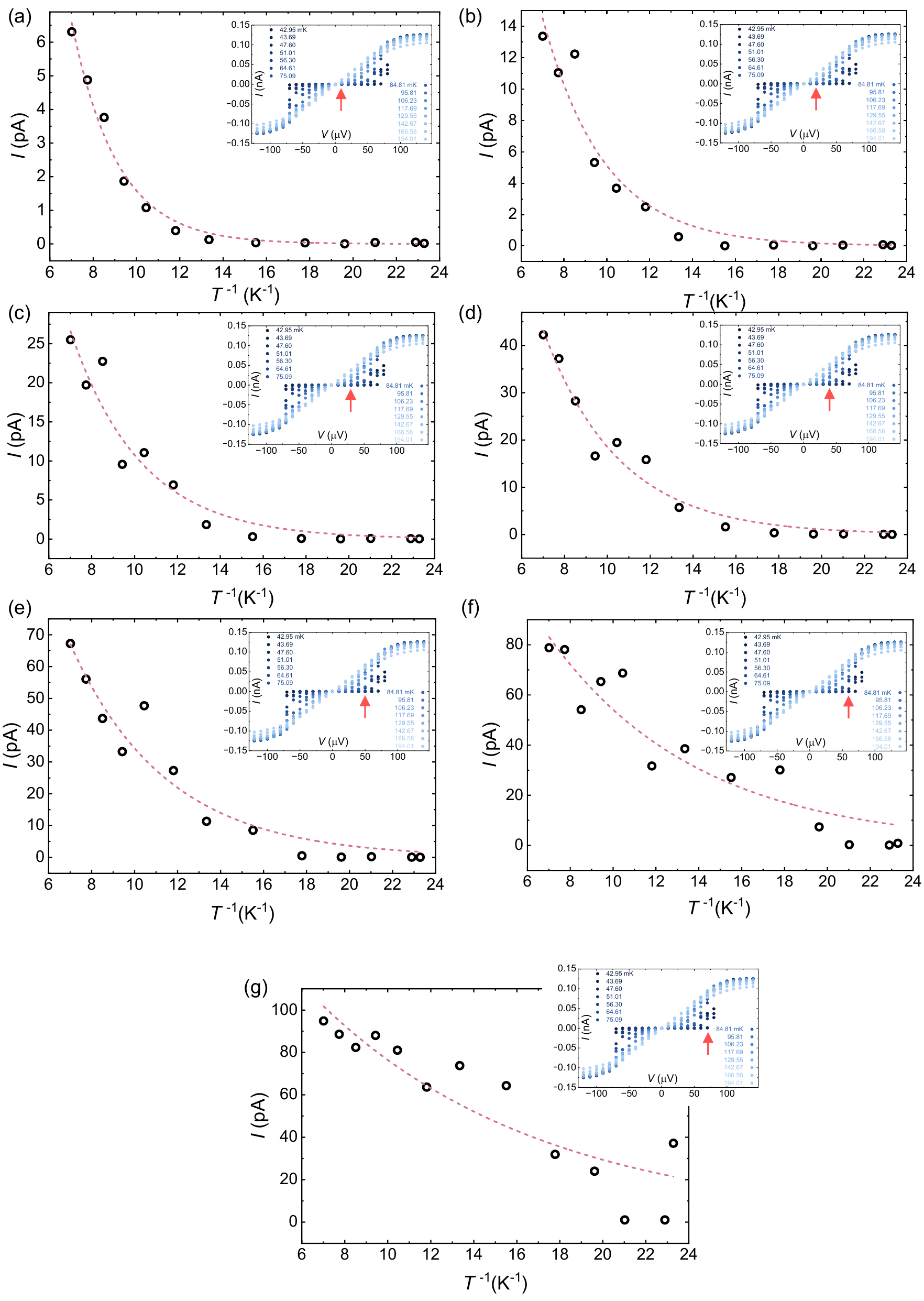}
	\caption{Arrhenius fits to the temperature dependence of the current measured at $V = 10\text{–}70~\mu$V in panels (a–g). In each panel, the inset indicates the corresponding voltage value with a red arrow. Symbols represent the experimental data, and dashed lines show fits to the Arrhenius law. The extracted values of the activation energy corresponding to the panels (a–g) are plotted in Fig.~3\,(c) in the main text.
		\label{I_vs_T-1}}
\end{figure}

\end{document}